\documentclass[11pt]{article}
\begin{document}
\title{Towards a New Democracy: Consensus Through Quantum Parliament}
\author{Diederik Aerts \\
        \normalsize\itshape
        Leo Apostel Centre for Interdisciplinary Studies (CLEA) \\
         \normalsize\itshape
         and Foundations of the Exact Sciences (FUND) \\
		\normalsize\itshape
		Department of Mathematics and Department of Psychology\\
		\normalsize\itshape
		Vrije Universiteit Brussel, 1160 Brussels, Belgium \\
        \normalsize
        E-Mail: \textsf{diraerts@vub.ac.be} \\
		}
\date{}
\maketitle
\begin{abstract}
\noindent
We compare different actual forms of democracy and analyse 
in which way they are variations of a ``natural consensus decision process''. We analyse how  ``consensus decision followed by majority voting'' is open to ``false play'' by the majority,
 and investigate how other types of false play appear in alternative 
types of  democratic decision procedures.
 We introduce the combined notion of ``quantum parliament'' and 
``quantum decision procedure'', and prove it
 to be the only one,  when applied after consensus decision, that is 
immune to false play.
\end{abstract}
\section{Introduction}
The aim of this article is to present a proposal for a new form of
democracy. We will give an explicit description of the structure of 
this new
democracy in the sections that follow. I decided to write a 
contribution on
this topic for the book ``Worldviews, Science and Us: Redemarcating
Knowledge and its Social and Ethical Implications'', because I was 
inspired
to elaborate this structure for a new form of democracy primordially 
while
reflecting on the nature of natural processes in the world. It was when
reflecting on the nature of ``quantum processes'' that I had a sudden
insight that brought several pieces of a puzzle together. More 
specifically,
it became clear to me how a possible remedy can be proposed for 
profound
shortcomings of the democratic process in our actual society. I found 
it to
be a good example of how reflections on one scientific discipline can 
lead
to fruitful insights in a seemingly completely different scientific 
field.
This proposal for a new form of democracy is, however, not a mere 
attempt to
apply insights into the nature of quantum processes to the political
processes in our society. It is just as firmly rooted in personal 
political
reflections generated by a long-felt concern about ``what is going on 
with
our western democratic systems''.

For years I have been formulating, both in private and in circles of 
close
friends, varying critiques of the functioning and practice of our 
democratic
system. I have, however, never made an effort to write down any of 
these
analyses, except for a short text that however remained in the form of 
a
preprint, and its English version \cite{aerts01,aerts02}. Finally, a 
variety
of reflections came together as pieces of a puzzle, and while I was
reflecting on the nature of quantum processes, made me see how a new 
form of
democracy could be presented that would be worthwhile considering as an
alternative to our current models.

\section{Democracies}

The word ``democracy'' originates from the Greek ``demos'' (``the 
people'')
and ``kratein'' (``to rule''). Hence the original meaning of democracy 
was:
``Rule by the People''. When referring to ``a democracy'', we therefore 
mean
a form of government in which ordinary citizens take part, in contrast 
with
a monarchy or dictatorship \cite{moore01,lijphart01}.

This article concentrates on the type of democracy that is now 
spreading all
over the world, and that emerged in West-Europe in the past centuries, 
with
roots in Ancient Greece \cite{levin01,dahl01}. This type of democracy is
often called ``representative democracy''. It comprises a form of 
government
in which voters choose, in free, secret and multi-party elections,
representatives to act in their interests. Globally, in 2004, a 
substantial
part of the world's people live in representative democracies, 
including
constitutional monarchies with a strong representative branch \cite{lijphart01}.

Let us call a representative democracy that works along the mechanism 
of
``majority voting'' a ``majority rule democracy''. In practice, this 
means
that a specific proposal that is debated in a nation's assembly of
representatives, \textit{e.g.} parliament, will be accepted if and only 
if
more than 50\% of such representatives vote in favour. If the 50\% is 
not
attained, the proposal is rejected. A higher percentage of votes in 
favour
is necessary, usually two thirds of the totality of votes, if the 
proposal
introduces a change in the constitution of the nation. This means that 
the
constitution plays the role of a more stable and less easy to change 
set of
rules.

\subsection{Majority rule democracy}

\begin{quote}
{\textquotedblleft The strongest is never strong enough to be always 
the
master, unless he transforms strength into right, and obedience into
duty\textquotedblright\ (Jean Jacques Rousseau in \cite{rousseau01})}
\end{quote}
We want to concentrate on the aspect of majority voting that takes 
place in
a majority rule democracy. Majority voting constitutes in effect a kind 
of
\textquotedblleft right of the strongest\textquotedblright\ in 
disguise. If
in principle the majority always gains, it is the biggest group, and 
hence
the strongest opinion, that always has to be followed after the voting 
has
taken place. One of our critiques of existing democracies is related to 
this
aspect of majority voting. We want to analyse many of its aspects, and 
hence
not only the immoral aspect, as suggested by the above quote from Jean
Jacques Rousseau \cite{rousseau01}. Rousseau argued that 
\textquotedblleft
the strongest is never strong enough to be always the
master\textquotedblright ,\cite{rousseau01} and hence that the
\textquotedblleft right of the majority\textquotedblright\ confers a 
much
more intrinsic power upon the strongest than is the case for the
\textquotedblleft strongest in nature\textquotedblright . Apart from 
this
ethical aspect, we will come to the conclusion that there are other 
aspects
that make majority rule democracy not the best candidate for a 
democracy.

It is commonly accepted that a majority rule democracy should not be
identified as an ethical form of democracy. A majority rule democracy 
is
defended usually for purely pragmatic and practical reasons: it is 
argued
that other types of democracy, the ones that do not adopt the majority 
rule,
are inefficient, because decision making takes too much time and energy 
of
the group of representatives. Let us consider the most important of 
these
other types of democracy, namely the ``consensus democracy''.

\subsection{Consensus democracy}

\label{consensusdemocracy} Consensus democracy is the application of
consensus to the process of legislation. Consensus is a process for 
group
decision-making. It is a method by which an entire group of people can 
come
to an agreement. The input and ideas of all participants are gathered 
and
synthesised to arrive at a final decision acceptable to all.

Consensus decision making is of a higher ethical standard, because it 
is
based on the principle that every voice is worth hearing, and that 
every
concern is justified. If a proposal makes any number of people, even if 
only
one person, deeply unhappy, it is considered that there is a valid 
reason
for that unhappiness, and that ignoring it might be a mistake. The 
pursuit
of consensus not only aims to achieve better solutions but also to 
foster a
sense of community and trust. With consensus, people can and should 
work
through differences and reach a mutually satisfactory position. It is
possible for one person's insights or strongly held beliefs to sway the
whole group. No ideas are lost, each member's input is valued as part 
of the
solution.

But, there are good reasons to be sceptic about a consensus democracy 
in
practice, because indeed it will often take a very long time to reach
agreement. This makes a consensus democracy, although ethically of a 
higher
standard than the majority rule democracy, not a very useful form of
democracy in practice. It often happens that new institutions start 
with the
implementation of a consensus democratic structure, because members can 
feel
safe then that they will not have to submit to a majority vote
decision whose outcome would be very bad for the people they represent. 
The
European Union is an example of an institution that makes use of such a
consensus decision structure. Usually, however, such institutions tend 
to
evolve steadily towards a majority rule democracy, the argument being 
that
the consensus democracy \textquotedblleft does not work in
practice\textquotedblright , and leads to too much inefficiency. In the 
case
of the European Union, the complaint is that many important decisions 
are
just not taken, because if one of the members does not agree the result 
is
that \textquotedblleft nothing happens\textquotedblright . The 
question:
\textquotedblleft Why does a consensus democratic system not work in
practice?\textquotedblright\ is one of the key issues that we will try 
to
analyse in this article.

\section{Natural and procedural decision processes}

If a group of friends decides to go for a walk in the nearby forest on 
a
sunny Sunday afternoon, their decisions have to be mutually agreed on, 
\textit{e.g.} the question ``will we stop for a drink in that pub along 
the
way?'', will most probably be resolved in some kind of ``natural way''. 
What
do we mean by ``natural''? We mean that no ``well defined procedure of 
how
to take decisions'' was agreed upon before the friends started out on 
their
walk. This is the way that most of the decisions that involve a group 
of
people are taken in our everyday world. We will call the decision 
process
that takes place in this way a ``natural decision process''. A decision
process that follows a well defined procedure we will call a 
``procedural
decision process''.

Sometimes, the distinction between a natural decision process and a
procedural decision process is not strict, for example, a process that
started as a natural decision process may well end up as a procedural 
one,
if it still fails to yield results after a reasonable amount of time 
has
elapsed. Suppose that during their walk the group of friends gets 
strongly
divided over the question of whether they will have a drink in the pub 
or
not, and that they cannot reach agreement including after discussing 
the
matter at some length; it could well be that one of them proposes to 
vote on
the issue, so that the natural decision process is changed into a 
procedural
one, at least, if all agree to this change. As a first remark, we 
should
note that for any type of procedural decision process going on in a 
group of
people it will at least be necessary to have a consensus about the 
procedure
to be followed in the procedural decision process amongst this group of
people. However, although this may not seem to be the case at first 
sight,
procedural decision processes have a deeply different structure as 
compared
to natural decision processes. To show this is the subject of the next
section.

\subsection{Boycott and false play in procedural decision systems}

Let us look in some more detail at the European Union. For many issues 
that
a majority of the member countries agreed upon, no decisions have been
taken, not even in an amended form, because some members, or indeed 
only a
single member nation, did not agree. This has happened on many 
occasions
within the \textquotedblleft procedural consensus decision
system\textquotedblright\ that the European Union adopts. We claim that 
this
phenomenon does not typify the \textquotedblleft natural decision
system\textquotedblright . If we return to the group of friends taking 
a
walk on a sunny Sunday afternoon, and if we suppose that no agreement 
is
reached on any of the decisions to be taken, it would be very plausible 
for
the group to decide to have a vote on the matter so as yet to find
agreement. But, even before they \textquotedblleft decide to have a
vote\textquotedblright , many other options might have been considered. 
In
any case, what is clear is that most probably the group of friends will 
not
allow their afternoon to be spoiled because a few or only one of them
disagrees about what the others want to do, threatening to boycott 
the
entire walk. The reason why such a boycott by a small minority, or one
person, rarely happens in everyday life, is because in everyday life no
fixed decision procedure has been agreed upon. Friends intending to 
have a
pleasant walk in the forest together will not decide beforehand on a
procedure to be followed in case they should stumble upon a 
disagreement
that cannot be resolved without a procedure. The natural decision 
process is
open to any kind of procedure at any moment, and it is exactly for this
reason that it cannot be reduced to a specific procedural decision 
process.
A procedural decision process that can be adapted at any moment and as 
often
as required is comparable to the natural decision process, but even 
such a
highly complex procedural decision process would be only an 
approximation of
the natural one.

\subsection{Pure consensus and majority consensus}

The European Commission follows the simplest consensus decision 
procedure of
all, namely ``if no consensus is reached no decision is taken''. Let us 
call
this the ``pure consensus system''. The weakness of this pure consensus
system is its vulnerability to boycotting by a small minority, which 
may
even be a single representative. Indeed, if the procedure of the pure
consensus system is known by every representative, it will be easy to
boycott the whole process by just ``not allowing consensus to be 
reached in
the time available''.

One of the possibilities to avoid such a boycott is to introduce a 
different
consensus procedure. In this procedure, a consensus is looked for 
initially,
but if it cannot be reached, the procedure of majority voting is 
followed.
Let us call this the ``majority consensus system''. This procedure is 
open
to false play too, however. If the majority consensus system is 
adopted, the
group of representatives will first look for a consensus, but if no
consensus is reached after a given time, which is fixed before the 
process
starts, they will change to the system of majority voting. No boycott 
of the
decision is yet possible, but false play is, obviously. Indeed, the 
majority
may well decide to prevent consensus from being reached, because they 
\textit{know} that this will be followed by a majority vote, so that 
they
will have the decision in the way they want it to be, without the need 
for
consensus.

This means that the majority consensus system is open to false play by 
the
majority, just as the pure consensus system is open to boycott by the
minority. Both procedures, pure consensus and majority consensus, are 
very
different from the natural decision process that we find around us in
everyday life. Can we find a procedural model that resembles more the
natural decision process?

\subsection{Random consensus}

\label{randomconsensus} Let us make the situation that we are 
considering
somewhat more concrete, such that we can look for alternative 
procedural
decision models. Suppose that an assembly of representatives consisting 
of $%
n $ people is gathering, where $n$ is sufficiently large, for example 
$100
\le n$ . They discuss a specific measure and different decisions in 
relation
with this measure are considered and proposed. Suppose that after a 
period
of discussion two alternative decisions are left, so that the assembly 
will
try to reach consensus considering both of them. However, no consensus
ensues in the period of time available. A majority of the 
representatives,
let us say $n-1$, is in favour of decision A, and one representative is 
in
favour of decision B. The two types of decision procedures that we have
considered so far would produce the following results. The pure 
consensus
system would result in ``no decision'' being made meaning that, in the
perceptions of the $n-1$ persons who are in favour of decision A, the 
one
person in favour of decision B has boycotted the overall process. The
majority consensus system, for its part, would result in decision A 
being
taken.

Suppose that a society using the majority consensus system has become 
aware
of (i) its unethical nature, and also, even more importantly, (ii) the
obvious possibility of false play and, as a consequence, of a
\textquotedblleft forced decision\textquotedblright , and suppose that 
it
wants to do something about it. More particularly, a way is 
investigated to
\textquotedblleft protect\textquotedblright\ the minority, in our 
example,
only one person out of $n$, who might, however, be representing a lot 
of
people. The following procedure could be proposed, which we will call 
the
\textquotedblleft random consensus system\textquotedblright . If no
consensus is reached after a well defined period, a random process is
organised to determine which of the alternatives will be chosen. In the 
case
of our specific example, this would come to tossing a coin and choosing 
for
decision A if the coin shows head and decision B if the coin shows 
tail.
Obviously, the minority gains by this random consensus system as 
compared
with the majority consensus system. In the case of our specific 
example, the
one representative gains a lot, because suddenly there is a 
fifty--fifty
chance of decision B or decision A being taken. But if all 
representatives
know in advance that this random consensus system is going to be 
applied, it
can be falsely used by the minority this time, in much the same way as 
the
majority consensus system can be falsely used by the majority. Indeed, 
the
minority, in our case the one representative who is in favour of 
decision B,
may well boycott the process of consensus, because he or she knows that
after the time for consensus has passed, the coin will be tossed, 
leaving
him or her with a 50 \% chance of his or her preferred decision being 
taken
in its pure form, instead of a consensus decision, which, given that $n-1$ representatives are in favour of decision A, will anyhow be closer 
to
decision A than to decision B. Our conclusion is that just as the 
majority
consensus system invites false play by the majority, preventing real
consensus, the random consensus system invites false play by the 
minority,
equally preventing real consensus.

\section{Quantum democracy}

In the introduction of this article I said that it was reflecting on 
the
nature of quantum processes that made me see how it would be possible 
to
propose a solution to some of the problems of our democratic system. 
Already
years ago, I used to give the example to my students of what I then 
called a
``quantum parliament''. Let me explain what such a quantum parliament 
is,
and why I found it an interesting idea at that time.

\subsection{Quantum parliament}

\label{quantumparliament} Suppose one considers a classical parliament, 
such
as the ones we know. This means that we have an overall group of
representatives constituting the parliament, and subgroups whose 
members
belong to different political parties. Let us be more concrete, and 
suppose
that we have a parliament of $n$ representatives, and that there are 
five
parties, which we will call A, B, C, D and E, where $n_{A}$, $n_{B}$, 
$n_{C}$%
, $n_{D}$ and $n_{E}$ are the number of representatives belonging to 
parties 
$A$, $B$, $C$, $D$ and $E$. This means that 
\begin{equation}
n_{A}+n_{B}+n_{C}+n_{D}+n_{E}=n  \label{sumrule}
\end{equation}%
Usually, a government is composed of a collection of parties such that 
the
sum of the representatives of all the composing parties is more than or
equal to $n/2$. As a consequence, whenever the parliament has to vote 
on a
certain proposition, the government can \textquotedblleft in some
way\textquotedblright\ obtain a majority vote for this proposition, and
hence have it decided the way the government wants. This indeed is the 
case
if all representatives of the parties that constitute the government 
follow
the government's opinion in their parliament vote. We stated
\textquotedblleft in some way\textquotedblright , because in principle 
this
does not have to be and even should not be the case. Indeed, in all 
western
democracies there is a strict division amongst the three powers: (1) 
the
executive power, in the hands of the administrative branch of the
government, including ministers, the cabinet, civil servants, the 
police and
the army; (2) the legislative power, in the hands of the lawmakers,
effectively the representatives of the parliament; and (3) the 
judiciary
power, more concretely the enforcers of the law, the judges, 
magistrates and
tribunals. But in practice, parliamentary decisions are often made by 
its
members that belong to the government. Apart from this, all parliaments 
in
western democracies decide through majority voting, which means that 
they
give rise to a majority rule democracy.

The quantum parliament follows a probability procedure, hence partly as
referred to in section \ref{randomconsensus}, but also different. The
probabilities are weighted by means of the sizes of the different 
decision
groups. More concretely, this means the following: we develop a random
machinery, such that the parties $A$, $B$, $C$, $D$ and $E$,
respectively, are attributed probabilities $p(A)={\frac{n_{A}}{n}}$, 
$p(B)={%
\frac{n_{B}}{n}}$, $p(C)={\frac{n_{C}}{n}}$, $p(D)={\frac{n_{D}}{n}}$ 
and $%
p(E)={\frac{n_{E}}{n}}$. From (\ref{sumrule}) it follows that 
\begin{equation}
p(A)+p(B)+p(C)+p(D)+p(E)=1
\end{equation}%
which means that we can interpret $p(A)$ (or $p(B)$, $p(C)$, $p(D)$, 
$p(E)$,
respectively) as the probability of party $A$ (or $B$, $C$, $D$, $E$,
respectively) deciding.

Hence the quantum parliament is different from a majority rule 
parliament,
because decisions are taken following a random procedure, which means 
that
also the smallest party can win, but it is also different from a random
consensus system, where each party would have an equal chance to win. 
For
each party, the chances to win are proportional to its size; hence the
bigger a party, the greater its chance to win.

\subsection{Quantum consensus}

Our proposal for a quantum consensus system is the following. Suppose a
specific proposal is made that requires a parliamentary vote. There is 
a
particular period of time available for seeking consensus, which is 
decided
on beforehand. After this time has run out, the quantum parliament is 
to
decide. Concretely, this means that a probabilistic procedure is 
carried out
such that the majority has a probability proportional to its size to 
win the
vote, and equally so the minority has a probability proportional to its 
size
to win the vote. In other words, although the majority has more chance 
to
win, the minority will always have a chance to win too, however small 
it may
be.

It is interesting to point out at this stage that this quantum 
consensus
system is not subject to ``false play'' and/or boycotting in the way 
that a
pure consensus system, a majority consensus system or a random 
consensus
system is. A pure consensus system can be easily boycotted by the 
minority,
since it will not be followed by a decision vote. The majority 
consensus
system is typically boycotted by the majority. They know that they just 
have
to wait for the voting to win with certainty. The random consensus 
system is
typically boycotted again on the initiative of the minority. Indeed, 
the
minority increases its power by waiting until a pure random decision is
made. A quantum consensus system is free from all these flaws. Indeed, 
if
the majority decided to boycott, they might lose the vote to the 
minority
because of the randomness of the procedure. The minority will not be 
tempted
to boycott the consensus either. Although they still have a chance to 
win if
they opt for boycotting the consensus and waiting for the vote, their 
chance
is definitely smaller than that of the majority, which makes this 
strategy
less attractive to them than seeking consensus.

The quantum consensus system is the only procedure that will stimulate 
both
sides, majority as well as minority, to strive for real consensus. It 
is the
only procedure that avoids that there is any benefit in boycotting the
consensus for either side, majority or minority. Nor does it entail the
disadvantage of the pure consensus procedure, namely that it takes too 
long
for a decision to follow. In the quantum consensus system, a fixed time 
is
reserved for consensus, after which the quantum parliament can decide in a
wink. This means that the quantum consensus 
system can
offer a real, efficient, workable and ethically balanced consensus 
decision
procedure.

\subsection{Natural and procedural, the aspect of determinism}

In this section we reflect about ``why there is a fundamental 
difference
between a natural and a procedural decision process'', ``why the 
quantum
consensus system is a good model for the natural decision process'', 
and
``what these insights tell us about the nature of processes in 
general''.

The types of boycotts and false plays that we mentioned in relation 
with the
different versions of procedural decision processes are only possible
because decision processes are instruments used by human beings, who 
have
the gift of foresight. Moreover, they are only possible because 
procedural
decision processes contain a definite deterministic aspect. For 
example, the
procedure of a procedural majority decision process is close to
deterministic, which means that once the majority and minority are 
known and
fixed, the outcome will be virtually known and fixed. Boycotting and 
false
play find their origin in this possibility of ``knowing the future''. 
It is
potential future events that influence the present through the minds of 
the
people involved in the process. Human minds manage to create a causal 
chain
from ``future potential'' to ``present actual''. Since any procedure 
that is
fixed and free from any randomness increases the potential of the human 
mind
to forecast the future, it also fortifies the causal link between 
future
potential and present actual. Once randomness is introduced in the
procedural process, the potential to forecast the future will decrease,
approaching the level of a natural decision process. This similarity is 
most
pronounced if the introduced randomness is quantum.

We know that causal effects of potential future to actual present also exist in the realm of the micro-world through the 
effect of
\textquotedblleft non-locality\textquotedblright . The question arises
whether the origin and structure of quantum randomness is not linked to 
the
function we pointed out in relation with decision processes, namely 
that the
quantum consensus system is the only procedural decision process that
demotivates the \textquotedblleft false play type\textquotedblright\ of 
use
of this causal \textquotedblleft future to present 
effect\textquotedblright
. There might be a Darwinian element of evolution involved that in the 
long
term makes the quantum process fitter than any other, and such that it 
was
selected in the course of time for the entities populating the 
micro-world.
Our analysis also indicates that the quantum formalism and more 
specifically
the quantum superposition state might deliver a good model for 
consensus as
a state within a process. We plan to investigate these questions in 
depth
within the approach that we put forward in \cite{aertsetal01,aertsetal02,aertsgabora01,gaboraaerts01}.

\end{document}